\documentclass[twocolumn,showpacs,preprintnumbers,amsmath,amssymb]{revtex4}

\usepackage{graphicx}
\usepackage{amsmath,amssymb}
\usepackage{dcolumn}
\usepackage{bm}
\usepackage{subfigure} 
\usepackage{natbib}
\usepackage[ansinew]{inputenc}

\begin{document}


\title{Bubble doubling route to strange nonchaotic attractor \\ in a
quasiperiodically forced Chua's circuit}

\date{\today}


\author{ K. Suresh} \email{sureshscience@gmail.com}
 \author{K. Thamilmaran}
\email{maran.cnld@gmail.com}
\affiliation{
Centre for Nonlinear Dynamics\\
School of Physics, Bharathidasan University\\
Tiruchirappalli - 620 024
}

\author{Awadhesh Prasad}
\email{awadhesh.prasad@gmail.com}
\affiliation{
Department of Physics \& Astrophysics\\
University of Delhi, Delhi - 110 007
}

\begin{abstract}
We have identified a novel mechanism for the birth of Strange Nonchaotic Attractor (SNA) in a quasiperiodically forced Chua's circuit. In this study the amplitude of one of the external driving forces is considered as the control parameter. By varying this control parameter, we find that bubbles appear in the strands of the torus.  These  bubbles start to double in number as the control parameter is increased. On increasing the parameter  continuously, successive doubling of the bubbles occurs, leading to  the birth of SNAs. We call this mechanism as the {\em bubble doubling} mechanism. The formation of SNA through this bubble doubling route is confirmed numerically, using Poincar\'e maps, maximal Lyapunov exponent and its variance and the distribution of finite-time Lyapunov exponents. Also a quantitative confirmation of the strange nonchaotic dynamics is carried out with the help of singular continuous spectrum analysis.
\end{abstract}

     
\maketitle

\section{\label{sec:level1}Introduction\protect\\}
Strange Nonchaotic Attractors (SNAs) are known to appear in quasiperiodically forced dynamical systems. They posseses a complicated geometrical structure, which shows them to be fractal in nature, but do not have any sensitive dependence on initial conditions, as seen from a negative  maximal Lyapunov exponent. Following the pioneering work by Grebogi {\em et al}. \cite{greb}, many researchers have found the existence of the SNAs, and have even classified the mechanisms for their occurrence in quasiperiodically forced continuous dynamical systems and maps. In particular the SNAs have been reported to arise in many physically relevant situations such as the quasiperiodically forced pendulum \cite{rome}, the quantum particles in quasiperiodic potentials \cite{bond}, biological oscillators \cite{ding}, the quasiperiodically driven Duffing-type oscillators \cite{venk,heag,yalc,kapi,feud}, velocity dependent oscillators \cite{venk1}, electronic circuits \cite{yang,zhiw,venk2}, and in certain maps \cite{pras,piko,piko1,kuzn,anis,nish,venk3,hunt,kim,lim,heag1}. Also, these exotic attractors were confirmed by an experiment consisting of a quasiperiodically forced, buckled, magnetoelastic ribbon \cite{ditt}, in analog simulations of a multistable potential \cite{zhou}, and in a neon glow discharge experiment \cite{ding1}. The SNAs are also related to the Anderson localization in the Schr\"odinger equation with a quasiperiodic potential \cite{keto,pras1} and they may have a practical application in secure communication \cite{zhou1,rama1,chac}.

Broadly the mechanisms for the birth of SNAs differ with the routes to SNA followed by the dynamical systems. A list of the various mechanisms pertaining to different routes to chaos are listed in Table - I \cite{pras2}.
Very recently, a new route namely the bubble route to SNAs have been reported by Senthilkumar {\em et al}. \cite{sent} in a quasiperiodically forced negative conductance series $LCR$ circuit with a diode, wherein, one of the driving force is a non-sinusoidal (square wave). In this route bubbles appear in the strands of the torus as a function of the control parameter. These bubbles then grow in size as the control parameter is increased. Subsequently the strands of the  bubbles start to wrinkle, giving birth to SNAs. The reason for this is that the quasiperiodic orbits become increasingly unstable in the transverse direction with the increase of the control parameter. This instability is induced by the square wave type quasiperiodic force resulting in an increase in the size of the doubled strands  (bubbles), followed by an enhanced wrinkling of these. 

\setlength{\tabcolsep}{3pt}
\setlength{\extrarowheight}{5pt}
\begin{table}
\caption{Routes and mechanisms for the formation of SNAs}
\small
\begin{tabular}{|p{4cm}|p{4cm}|}

\hline
{\bf \emph \;\;\;\;\;\;\;\;\;\;\;\;Type of route} & {\bf \emph \;\;\;\;\;\;\;\;\;\;\;\;\;\;Mechanism}  \\
\hline
Heagy-Hammel \cite{heag1}  & Collision of period-doubled torus with its unstable
parent\\
\hline
Gradual Fractilization \cite{nish} & Increased wrinkling of torus without any interaction
with nearby periodic orbits\\\hline
On-off intermittency \cite{yalc} &Loss of transverse stability of torus\\
\hline
Type-I intermittency \cite{pras} & Due to saddle-node bifurcation, a torus
is replaced by SNA\\
\hline
Type-III intermittency \cite{venk2}  & Subharmonic instability\\
\hline
Homoclinic collision \cite{pras1} & Homoclinic collisions of the
quasiperiodic orbits\\
\hline
\end{tabular}
\end{table} 
In this present work, we report successively doubling of the bubbles route to SNAs in a quasiperiodically forced Chua's circuit. Here, the bubbles appear in the strands of the torus as a function of the control parameter, then grow in size and get doubled in number. After many successive doubling of the bubbles, SNAs are born. All through these, the remaining part of the strands of the torus which do not contain the bubbles, remain  unaffected even though varying the control parameter. In this route no wrinkling of the bubbles are found to occur \cite{sent}. In order to confirm this, we present here a detailed numerical analysis using Poincar\'e map, Fourier spectrum, maximal Lyapunov exponent, its variance, finite time Lyapunov exponents and separation of near by trajectories in phase space. Further the singular continuous spectrum analysis is used to find scaling the exponent and fractal paths.

The paper is organized as follows. In Sec. II, we discuss the circuit realization of the quasiperiodically forced Chua's circuit. In Sec. III, the observations of the bubble doubling route to SNAs are given with the help of Poincar\'e cross section. In Sec. IV, we substantiate the observations by the maximal Lyapunov exponent spectrum and its variance. In Sec. V, we find the scaling exponent and fractal path by using singular continuous spectrum analysis. In Sec. VI, we discuss another qualitative measure of this route by separation of near by trajectories. In Sec. VII, the finite time Lyapunov exponents are used to confirm the SNA and show the mechanism of the bubble doubling route. Finally, in Sec. VIII, we summarize our results.\\
\section{\label{sec:level2}Circuit Realization}
The quasiperiodically forced Chua's circuit \cite{zhiw} is shown in Fig. (1). Using Kirchoff's laws, the equations of this circuit derived are\\
\begin{subequations}
\begin{eqnarray}
C_1\frac{dv_{C_{1}}}{dt}&=&G(v_{C_{2}}-v_{C_{1}})-g(v_{C_{1}}),\\
C_2\frac{dv_{C_{2}}}{dt}&=&G(v_{C_{1}}-v_{C_{2}})+i_{L},\\
L\frac{di_L}{dt}&=&-v_{C_{2}}+F_1\sin(\omega_{1}t)+F_{2}\sin(\omega_{2}t).\\ \nonumber
\end{eqnarray}
Here,
\small
\begin{eqnarray}
\hspace{-1.0 mm}g(v_{C_1})=\hspace{-1.0 mm}G_bv_{C_{1}}+0.5(G_a-G_b)[|v_{C_{1}}+B_p|-|v_{C_{1}}-B_p|]\\ \nonumber
\end{eqnarray}
\end{subequations}
\begin{figure}
\includegraphics[width=0.8 \columnwidth]{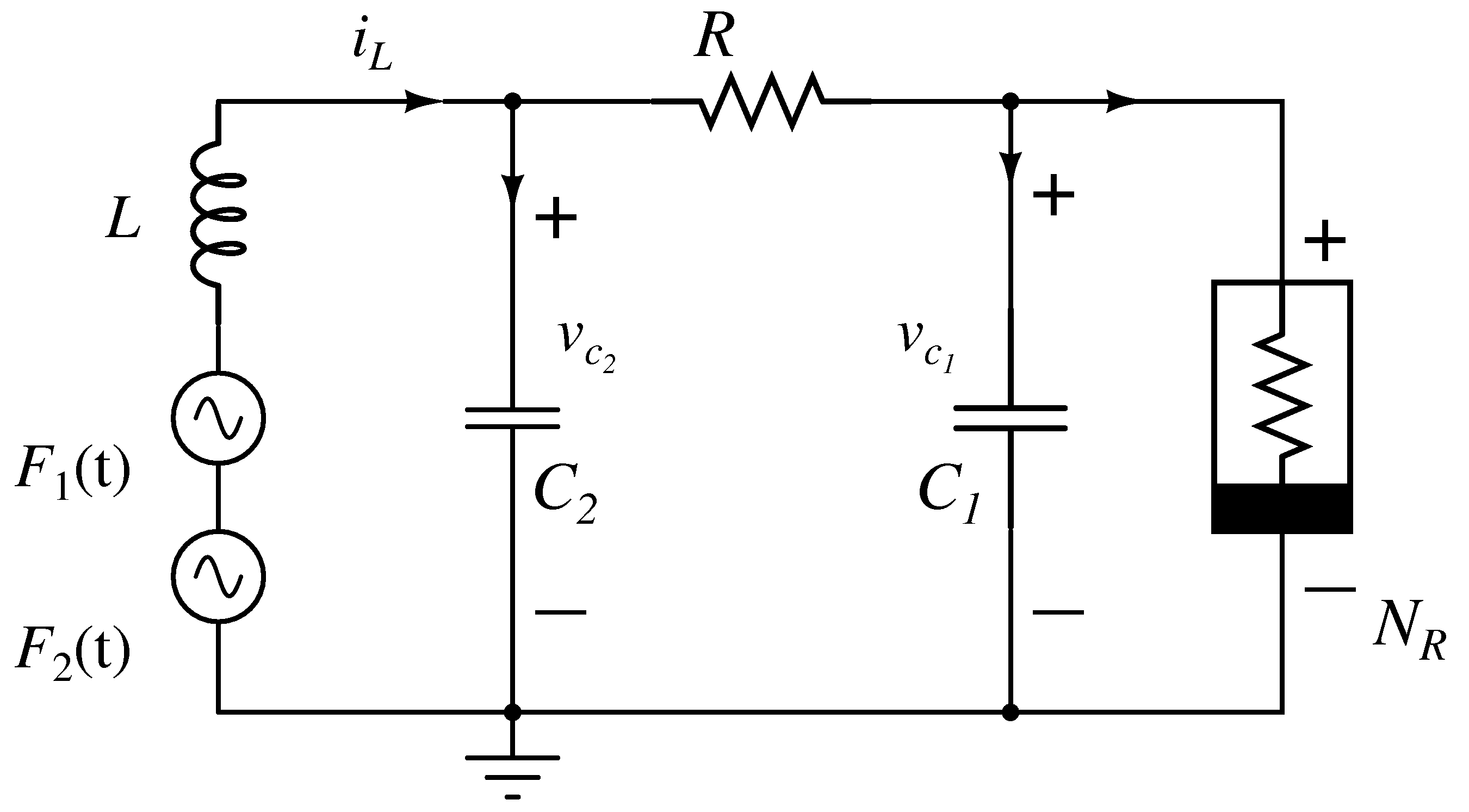}
\caption{\label Circuit realization of the quasiperiodically forced Chua's circuit. Here $N_R$ is the Chua's diode. The parameter values of the circuit elements are fixed as $C_1=16.8$ nF, $C_2=98.0$ nF, $L=18.75$ mH, $R=1565$ \ensuremath{\Omega} , $G_{a}= - 0.758$ mS,  $G_{b}=-0.409$ mS, ${\nu_1}  = 2849.0$ Hz, ${\nu_2} = 3560.0$ Hz and $F_2 = 5$ mV are chosen respectively. The forcing amplitude $F_1$ is taken as control parameter which being varied in our analysis.}
\end{figure}\\
where $v_{C_{1}}$, $v_{C_{2}}$  are the voltages across the capacitors  $C_1$  and $C_2$, and $i_L$  is the current through inductor $L$ respectively. $F_1 \sin(\omega_1t)$ and $F_2 \sin(\omega_2t)$ are the external excitations  having incommensurate frequencies, $g(v_{C_{1}})$ is mathematical expression of Chua's diode. where $G_b$ is the outer slope and $G_a$ is the inner slope of the nonlinear  curve. $B_P$ is the break point voltage.  The parameter values of the circuit elements are fixed as $C_1=16.8$ nF, $C_2=98.0$ nF, $L=18.75$ mH, $R=1565$ $\Omega $, $G_{a}= - 0.758$ mS and $G_{b}=-0.409$ mS. The value of the $\nu_1  (=\omega_1/2 \pi)$ and $\nu_2  (=\omega_2/2 \pi)$ are chosen as $2849.0$ Hz and $3560.0$ Hz respectively. The forcing amplitude $F_2$ is fixed as 5 mV and the  forcing amplitude $F_1$ is assumed as the control parameter. 
In order to study the dynamics of the circuit, the circuit Eqs. (1) can be normalized using the following rescale variables and parameters. $v_{C_{1}}$ $=$ $xB_p$, $v_{C_{2}}$ $=$ $yB_p$, $i_L$ $=$ $B_pGz$, $t$ $=$ $C_2\tau/G$, $G$ $=$ $1/R$, $a$ $=$ $RG_{a}$, $b$ $=$ $RG_{b}$, $\alpha$ $=$ $C_2/C_1$, $ \beta$ $=$ $C_2R^{2}/L$, $f_1$ $=$ $F_1\beta/B_p$, $f_2$ $=$ $F_2\beta/B_p$, $\omega_1$ $=$ $G\Omega_1/C_2$ and $\omega_2$ $=$ $G\Omega_2/C_2$.
\begin{subequations} 
\begin{eqnarray} 
\frac{dx}{dt}&=&\alpha(y-x-h(x)),\\
\frac{dy}{dt}&=&x-y+z,\\
\frac{dz}{dt}&=&-\beta y+f_1\sin(\Omega_{1}t)+f_{2}\sin(\Omega_{2}t),\\ \nonumber
\end{eqnarray}
where,
\begin{eqnarray} 
h(x)=bx+0.5(a-b)[|x+1|-|x-1|].\\ \nonumber
\end{eqnarray} 
\end{subequations} 
Eqs. (2) are numerically integrated using Runge - Kutta fourth order algorithm with step size 0.0001. The values of the normalized parameters are $a$ $=$ $-1.1862$, $b$ $=$ $-0.6400$, $\alpha$ $=$ $5.8333$, $\beta$ $=$ $12.8012$, $\Omega_1$ $=$ $ 2.7454$, $\Omega_2$ $=$ $3.4306$ and $f_2$ $=$ $ 0.1280$. The normalized parameter  $f_1$ of external quasiperiodic forcing is taken as control parameter.
\section{\label{sec:level2}Bubble Doubling Route to SNA}
To observe the emergence of SNA the amplitude of the external forcing $f_1$ is varied in the range $f_1\in(0.7296 , 0.8500)$. The Poincar\'e surface of section plot of the two strands corresponding to period$-$2 torus for the value of $f_1$ $=$ $0.7296$ is shown in Fig. 2(a) and the corresponding power spectrum are depicted in Fig. 3a(ii). As the value of the parameter $f_1$ is increased further, bubbles appear in the strands of the torus, for a critical value $f_1$ $=$ $0.7744$ these are shown in Fig. 2(b). 
\begin{figure}
\includegraphics[width=1.0\columnwidth]{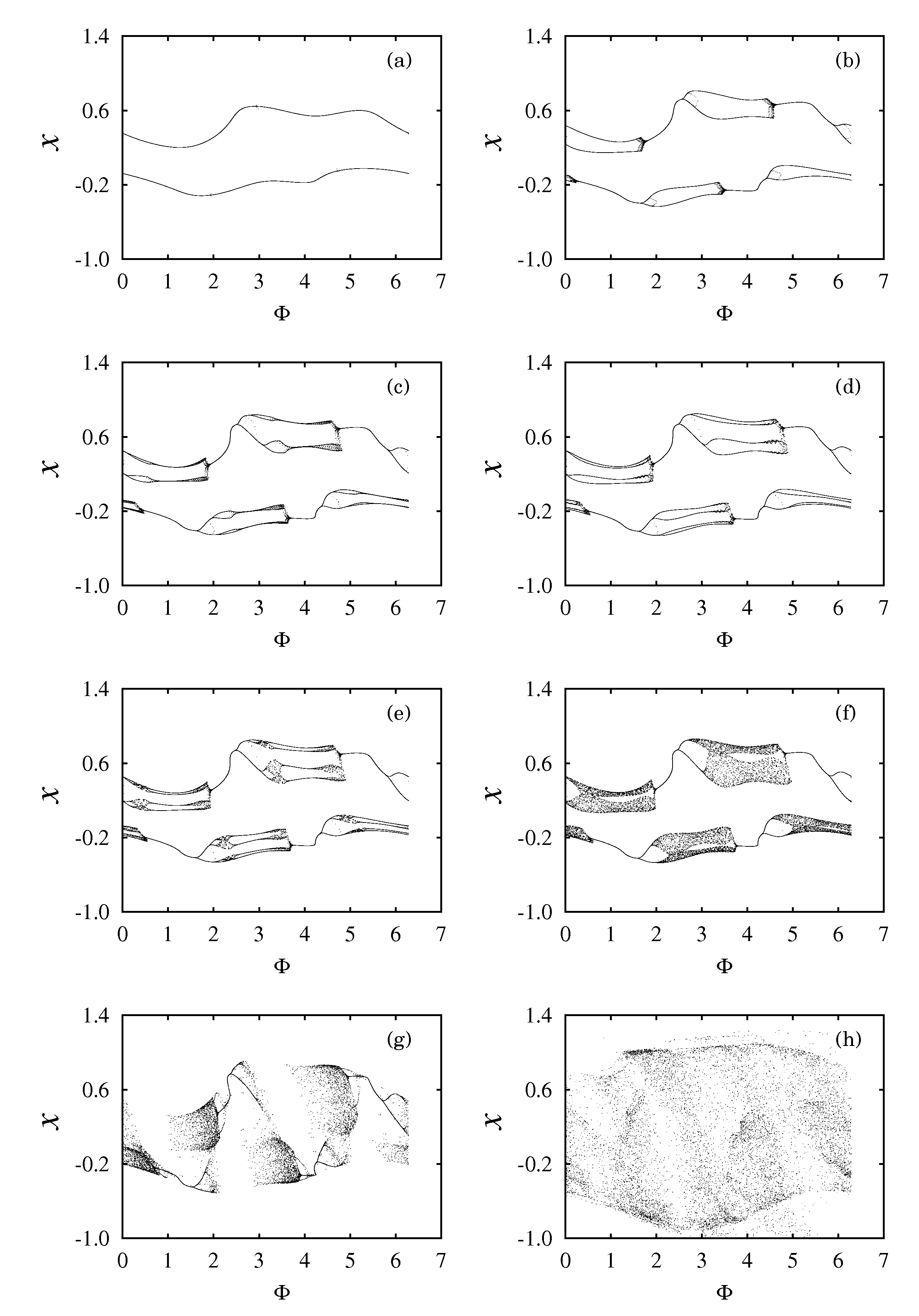}
\caption{\label{}Projection of the numerically simulated Poincar\'e surface section plots of the attractors of Eqs. (2) in the ($\phi - x$) plane with $\phi$ modulo 2$\pi$ as a function of amplitude of the forcing $f_1$ including the transition from quasiperiodic attractors to SNA through bubble doubling route: (a) period$-$2 torus for $f_1$ $=$ $0.7296$, (b) bubbled strands of period$-$2 torus $f_1$ $=$ $0.7744$, (c \& d) doubled bubble strands of the period$-$2 torus for $f_1$ $=$ $0.7770$ and $f_1$ $=$ $0.7803$,  (e) bifurcated bubble, (f) SNA for $f_1$ $=$ $0.8200$, (g) merged two torus and (h) chaotic attractor for $f_1$ $=$ $0.9475$.}
\end{figure}
Further, increasing the $f_1$ value from $0.7744$ the bubbles get doubled as shown in Figs. 2(c \& d). Beyond the value of $f_1$ $=$ $0.7803$, the bubbles get successive doubling Fig. 2(e) and  cause the emergence of SNA. The SNA start to appears above the value of $f_1$ $=$ $0.8123$. It is to be noted that the remaining part of the strands of the torus remain unaffected. The Poincar\'e surface section plot of SNA for the value of $f_1$ $=$ $0.8200$ is shown in Fig. 2(f) and the corresponding  power spectrum is shown in Fig. 3b(ii). Beyond the value of $f_1$$=$ $0.8200$, the bifurcated strands corresponding to period$-$2 torus merge into a single strand and then transit to chaotic attractor, which are shown in Fig. 2(g) and Fig. 2(h) respectively. The power spectrum of the chaotic attractors is shown in Fig. 3c(ii). These observations show clearly that  the bubbles that appear in the strands of the torus undergo successive doubling and cause the emergence of strange nonchaotic dynamics. The mechanism for this route is that the quasiperiodic orbit becomes unstable as a function of control parameter and resulting the appearance of bubbles in certain parts of the main strands. Further increasing the control parameter these bubbles losses it's stability and gets successive doubling and leads to birth of SNAs. These observations point out that this route is significantly  different from other routes known in dynamical systems \cite{pras2}. 
\\
\begin{figure}
\includegraphics[width=1.1\columnwidth]{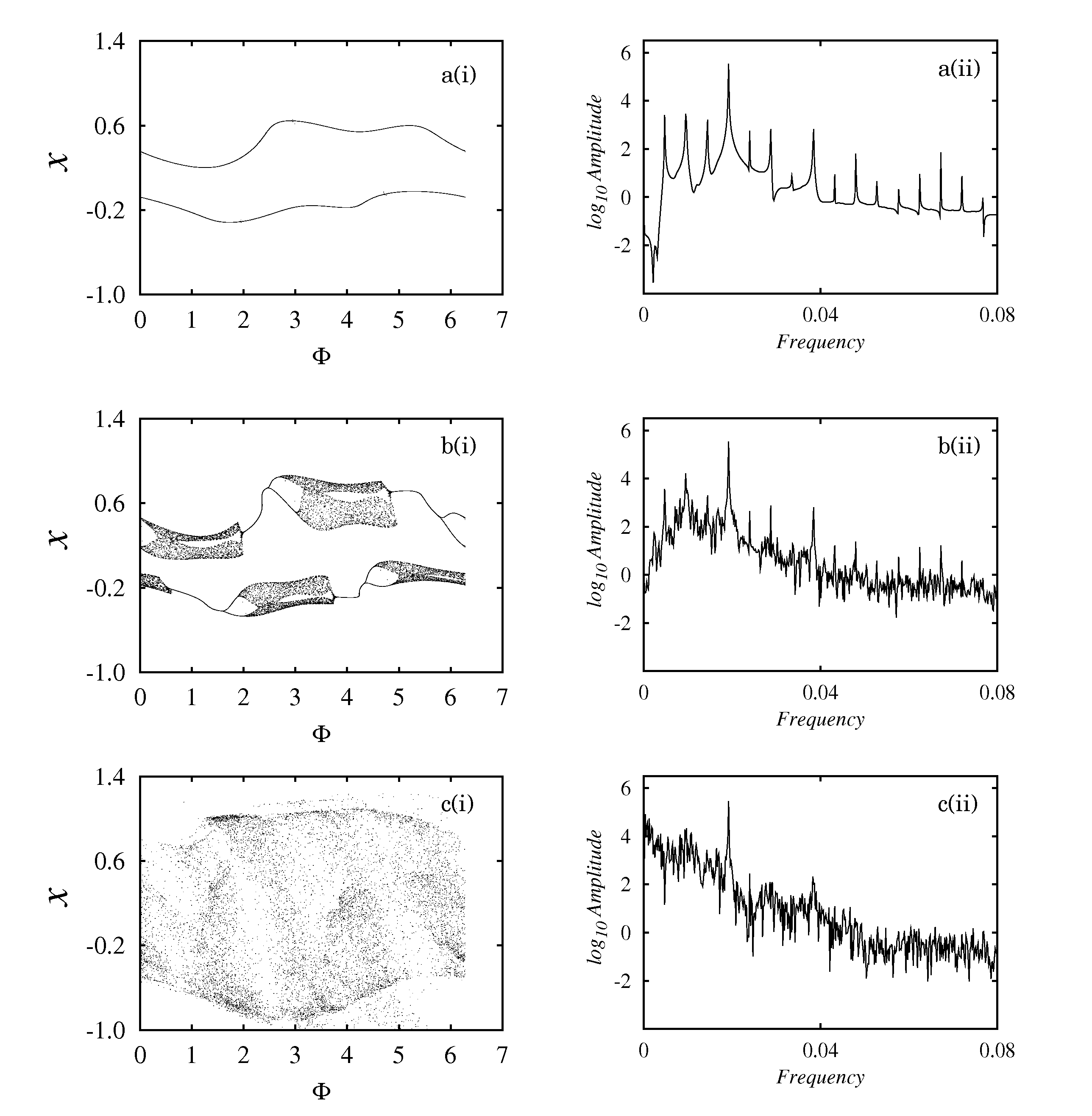}
\caption{\label{fig:epsart}Projection of the numerically simulated attractors of Eqs. (2) with  (i) Poincar\'e plot with $\phi$ modulo 2$\pi$ in the ($\phi - x$) plane and (ii) corresponding power spectrum for various value of $f_1$ indicating the transition from a quasiperiodic attractor to an SNA and chaos. (a) period$-$2 torus for $f_1$ $=$ $0.7296$, (b) SNA for $f_1$ $=$ $0.8200$ and (c) chaos for $f_1$ $=$ $0.9475$.}
\end{figure}
\section{\label{sec:level2}Lyapunov Exponent and its variance}
To characterize the quasiperiodic, strange nonchaotic and chaotic dynamics, the largest Lyapunov exponent and its variance are calculated as functions of $f_{1}$ for the range of 0.8000 to 0.8500.  The variance$(\sigma)$ of the largest asymptotic Lyapunov exponent $(\Lambda)$ from finite time Lyapunov exponents $(\bar{\lambda})$ of length M, and its variance defined as
\begin{subequations}
\begin{eqnarray}
\bar{\lambda}&=&\frac{1}{M}\sum_{i=1}^{M}\lambda_i,\\
\sigma&=&\frac{1}{K}\sum_{i=1}^{K} (\Lambda-\bar{\lambda_i})^2.
\end{eqnarray}
\end{subequations}
\begin{figure}
\includegraphics[width=1.0\columnwidth]{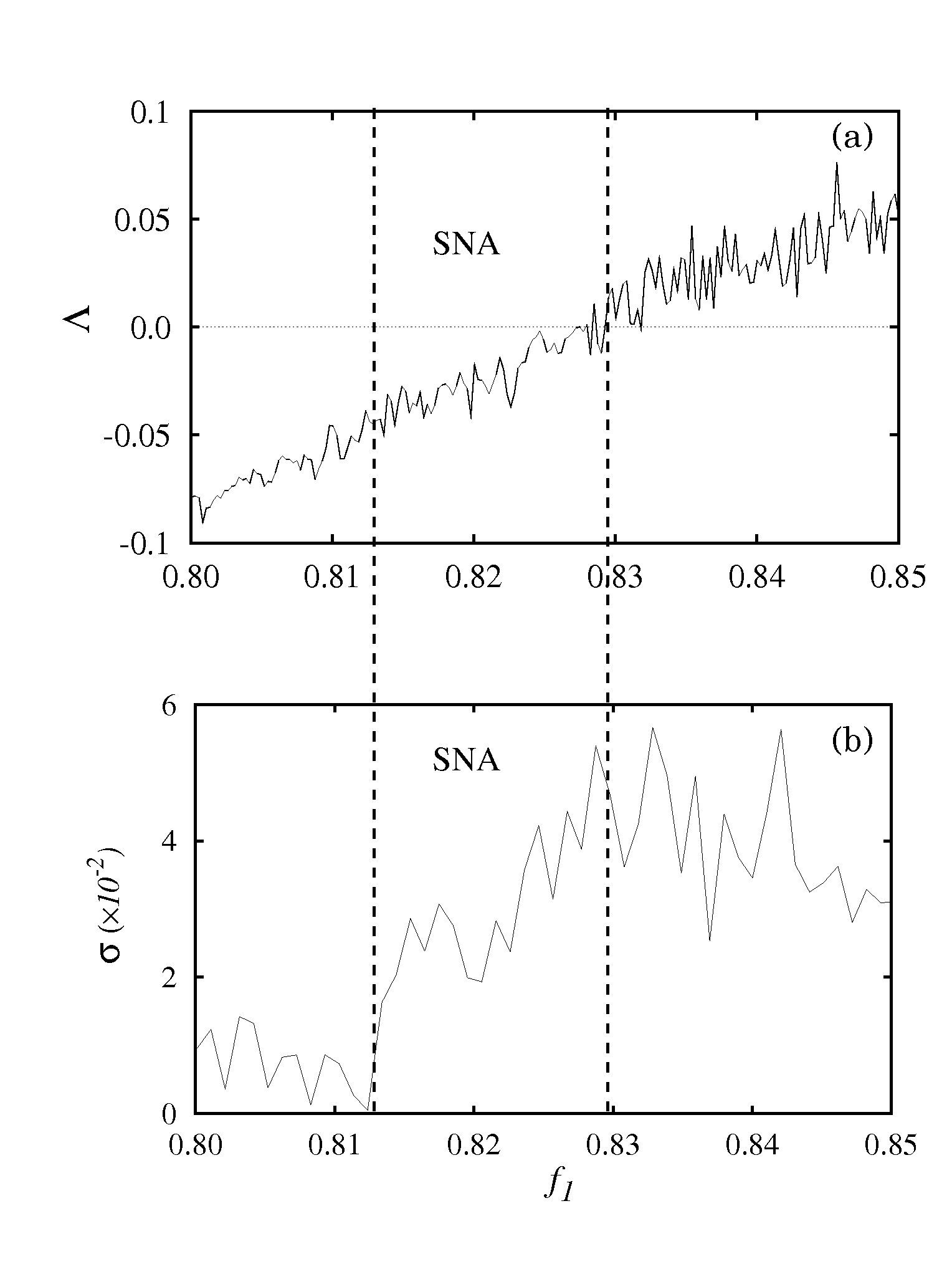}
\caption{\label{fig:epsart}Transition from two torus to chaos through SNA. (a) the largest Lyapunov exponent \ensuremath{\Lambda} and (b) variance \ensuremath{\sigma} as a function of parameter $f_1$. }
\end{figure}

where $K$ is the total number of finite LE (${\bar{\lambda}}$) and $\lambda_i$ is the instantaneous Lyapunov exponent. The plot of the largest Lyapunov exponent and its variance are shown in Figs. 4(a \& b).  It clearly demarcate the range of the three dynamical behaviours such as quasiperiodic, SNA and chaos. Below the value of $f_{1}$ $=$ $0.8123$ the system exhibits quasiperiodic behavior. In the range $f_{1}$ $=$ $0.8123$ to $f_{1}$ $=$ $0.8292$ the largest Lyapunov exponent is negative and the corresponding variance is higher in this range, is indicate the existence of SNA. The strange nonchaotic dynamics becomes chaotic dynamics beyond the value of $f_{1}$=0.8292 (after the largest Lyapunov exponent crosses zero axis).
\section{\label{sec:level2}Singular Continuous spectrum analysis}
The singular continuous spectrum analysis, which is used to quantitatively confirm the strange nonchaotic nature of the dynamics,  was first proposed in the investigation of model of quasiperiodic lattices and quasiperiodically forced quantum systems \cite{aubr}. In general, power spectra of dissipative dynamical system can be either discrete, or continuous, or a combination of both. Discrete spectrum are usually generated by regular motions such as periodic or quasiperiodic motions, where as continuous spectra correspond to irregular motions such as chaotic or random motion. A singular continuous spectrum is a mixture of both discrete and continuous spectra \cite{yalc}. Using this property, we can identify whether the bubble doubling indeed leads to strange nonchaotic dynamics or not. To confirm this we compute the partial Fourier sum \cite{piko1,yalc1} as
\begin{equation}
X(\alpha,N)=\sum_{k=1}^{N}(x)_{k}\exp(2\pi ik\alpha)
\end{equation}
where $\alpha$ is proportional to the irrational driving frequency $\Omega_{1}$ and $\{x_k\}$  is the time series of the variable $x$ of length N for the value of $f_1$ $=$ $0.8200$. When N is regarded as time, $|X(\alpha,N)|^2$ grows with time N as \cite{piko2} 
\begin{eqnarray}
|X(\alpha,N)|^2&\sim&N^\mu 
\end{eqnarray}
where $\mu$ is the scaling exponent. The time evolution  of $X(\alpha,N)$ can be represented by an orbit or a walker in the complex plane $(Re[X(\alpha,N)],Im[X(\alpha,N)])$ and for a singular continuous spectrum (when the dynamics is strange) it implies that the walk on the plane $(Re[X(\alpha,N)], Im[X(\alpha,N)])$ will be a fractal  \cite{piko1}.
\begin{figure}
\includegraphics[width=1.0\columnwidth]{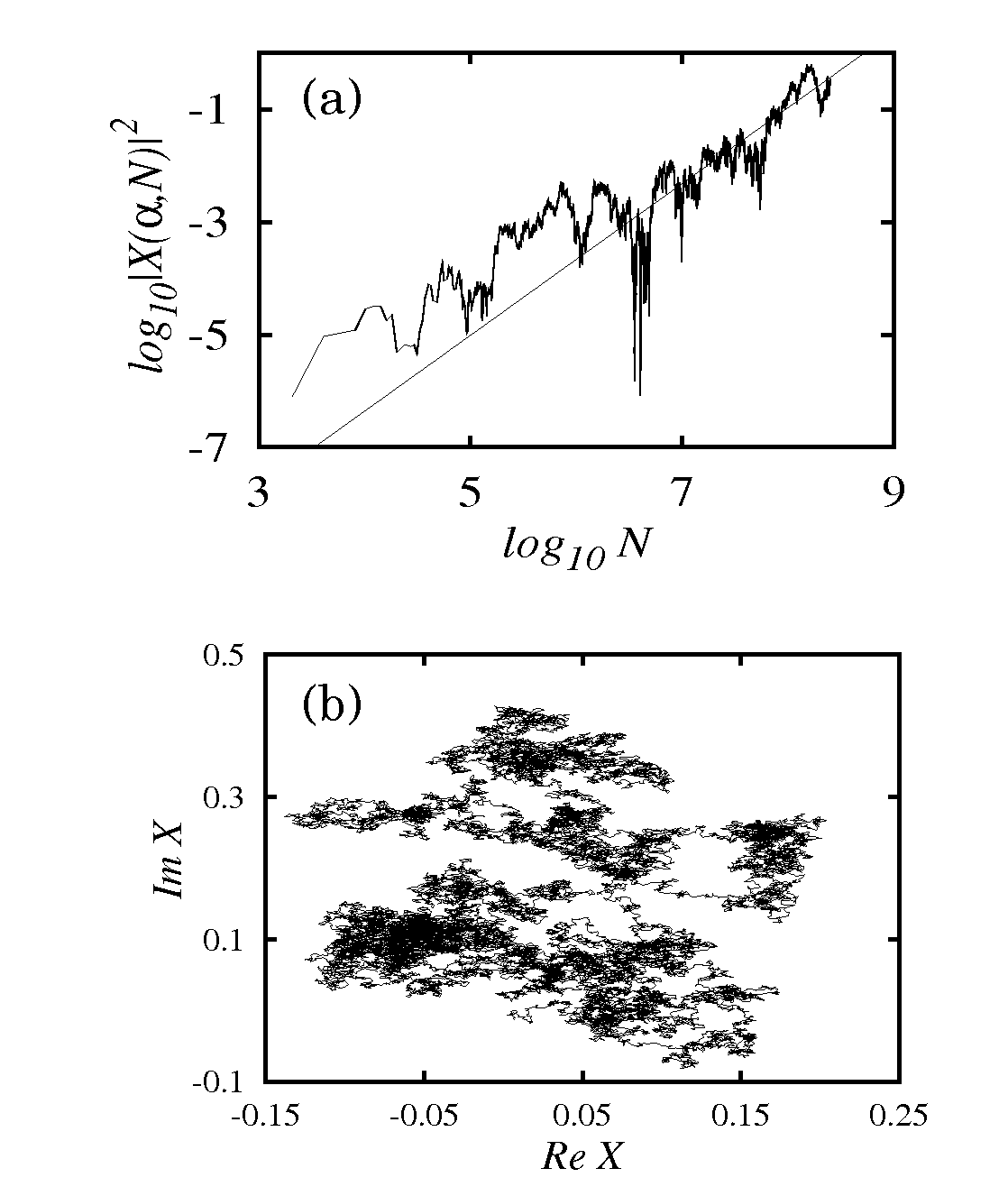}
\caption{\label{fig:epsart} Singular continuous spectrum analysis of the $\{x_k\}$ time series at  $f_1$ $=$ $0.8200$.
(a) plot of $N$  $vs$ $|X(\alpha,N)|^2$ showing the power-law scaling; the slop is $ \approx $ 1.345 and 
(b) the fractal path in the complex plane $(Re[X(\alpha,N)], Im[X(\alpha,N)])$}
\end{figure}
Fig. 5(a) shows, the plot of $N$ $vs$  $|X(\alpha,N)|^2$ for $\alpha$ $=$ $\Omega_{1}$/4, which has the scaling exponent $\mu$ $\approx$ 1.345 and the walk in Fig. 5(b) appears to be fractal. These results, $1<\mu<2$ and fractal walk,
strongly suggest that the dynamics is indeed strange and nonchaotic. 
\section{\label{sec:level2}separation of near by points}
SNAs exhibit complicated geometrical structure like chaotic attractors. One way to distinguish SNAs from chaotic attractors is to look for the sensitive dependence on initial conditions \cite{venk1}. In order to verify this sensitive dependence, we analyse the separation $\Delta$ between two orbits starting from two near by initial conditions. For this two near by points on the attractor ($x_{i}, y_{i}, z_{i}$) and ($x_{j}, y_{j}, z_{j}$) are chosen and their separation\\ 
\begin{equation}
\small
\Delta = \sqrt{(x_{i}-x_{j})^2+(y_{i}-y_{j})^2+(z_{i}-z_{j})^2}
\end{equation}
\begin{figure}
\includegraphics[width=1.0\columnwidth]{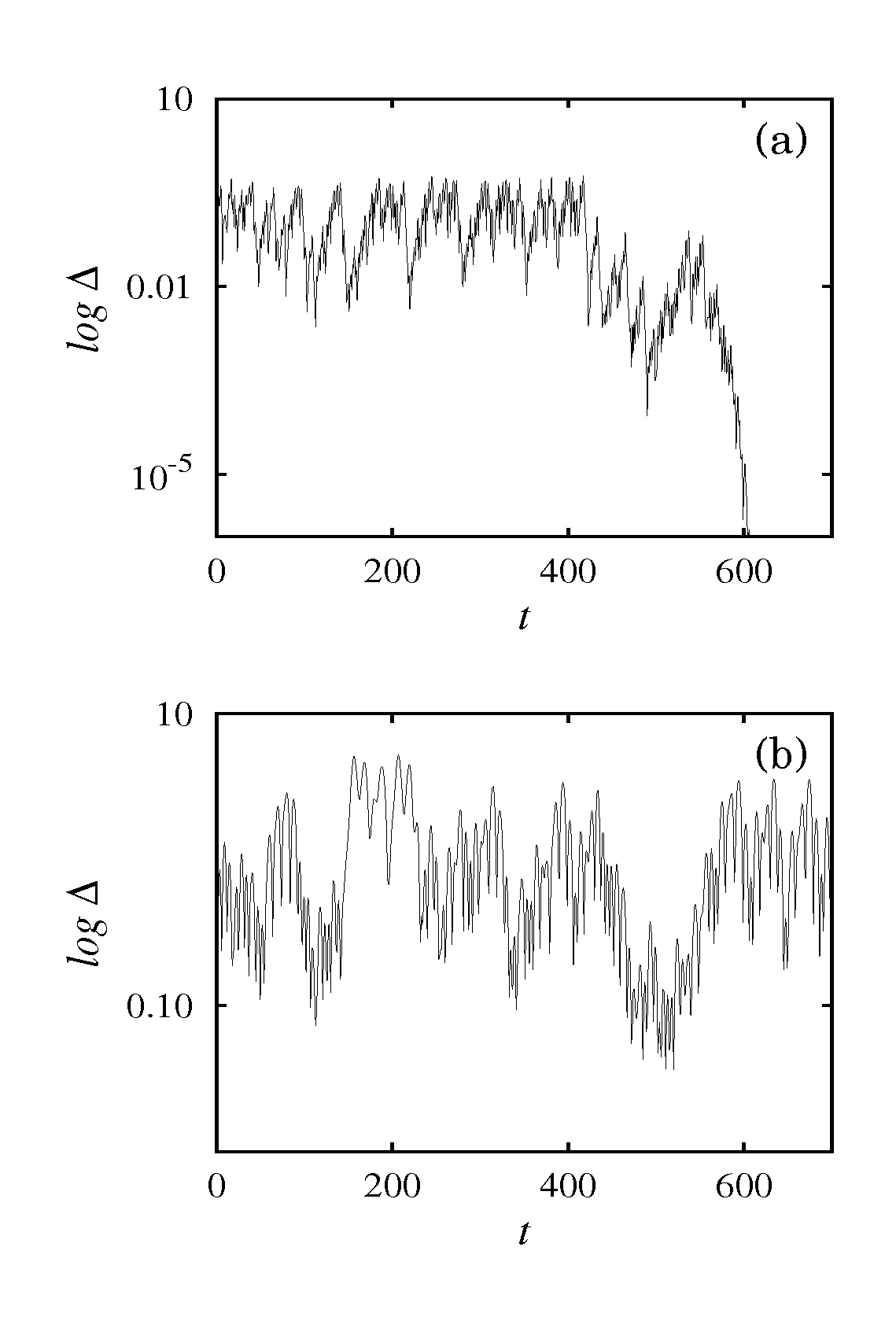}
\caption{\label{fig:epsart}Separation of two nearby points vs time (t). (a) strange nonchaotic
attractor for the value of $f_1$ $=$ $0.8200$ and (b) chaotic attractor for the value of $f_1$ $=$ $0.9475$.}
\end{figure}

is monitored at each forward step. For regular behaviour $\Delta$ will decay to zero as t$\rightarrow\infty$, but for chaotic behaviour, $\Delta$ becomes irregular. In Figs. (6) a plots of t versus $\Delta$ is shown. In  Fig. 6(a) shows  $\Delta$ for the value of $f_{1}$ $=$ $0.8200$ diminishes to zero in short interval of t and Fig. 6(b) shows  $\Delta$ for $f_{1}$ $=$ $0.9475$ sustains the irregular variation for long time (t$\rightarrow\infty$). Hence the former case clearly supports the loss of sensitive dependence on initial condition, which corresponds to the strange nonchaotic attractor, while the later one corresponds to a chaotic attractor. 
\section{\label{sec:level2}Distribution of finite time Lyapunov exponent}
We have discussed about the bubble doubling dynamics qualitatively in section III, by using Poincar\'e surface of section plots in the $(\phi-x)$ plane, also quantitatively by using maximal Lyapunov exponent and sigular continuous spectrum analysis in section IV. In addition to the above it is also possible to distinguish torus and SNA using distribution of finite time Lyapunov exponent \cite{mani}.
\begin{figure}
\includegraphics[width=1.0\columnwidth]{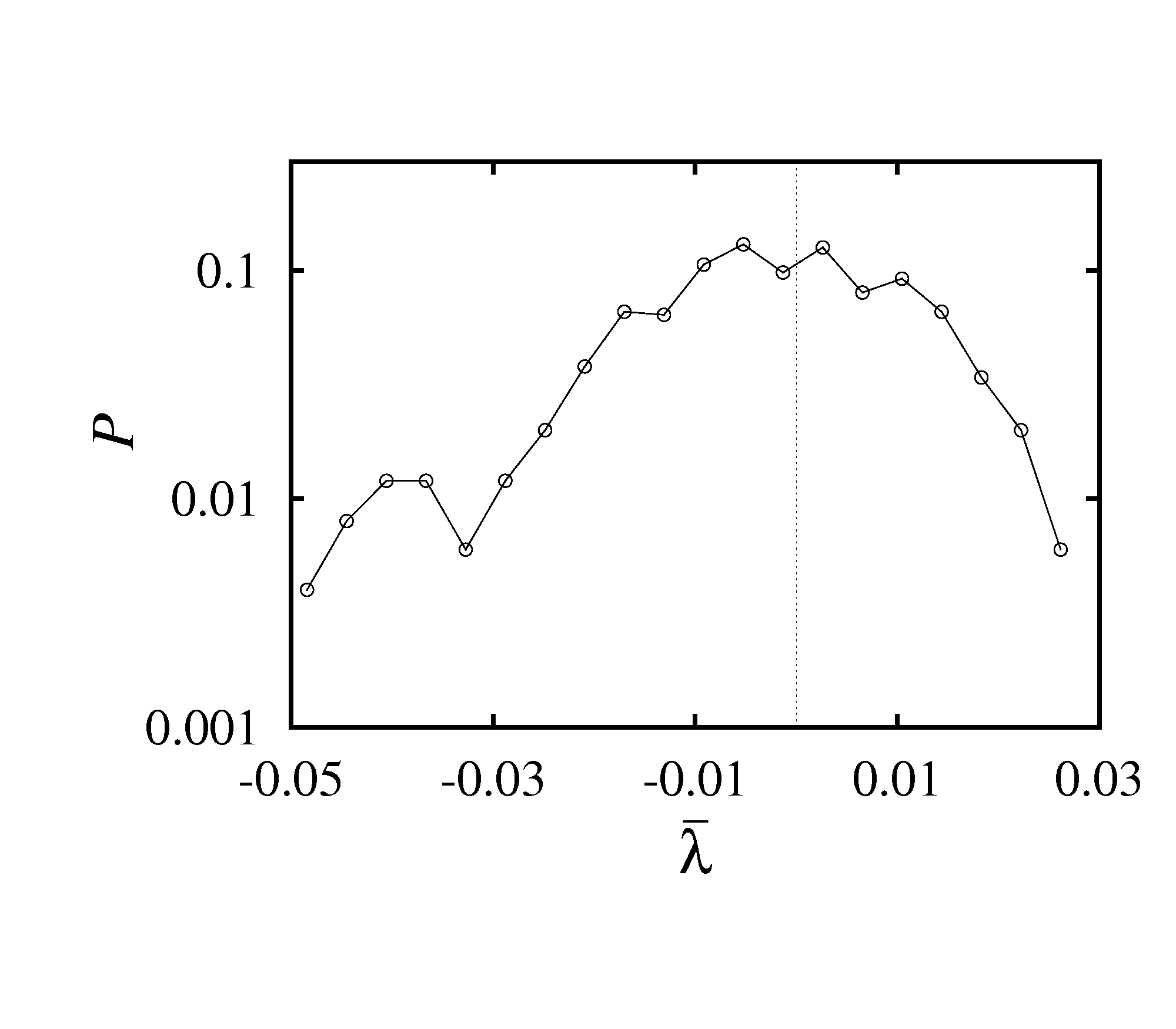}
\caption{\label{fig:epsart}Distribution of finite time Lyapunov exponent is taken for $M=2000$ for the value of $f_1$ $=$ $0.8200$}
\end{figure}
It has been shown \cite{pras} that a trajectory on a SNA actually possesses positive Lyapunov exponent in finite time intervals, although the asymptotic exponent is negative. As a consequence, it is possible to observe different characteristic of the  SNA created through different mechanisms \cite{sent} by studying the differences in the probability ($P$) of distribution of finite time Lyapunov exponent $P(M, \bar{\lambda})$ for positive and negative values.  We have calculated the distribution of finite time Lyapunov exponent $P(2000, \bar{\lambda})$ for the attractor shown in Fig. 3b(i) and have plotted the same in Fig. (7). Here we find that the distribution  exhibits an elongated tail for the negative values, thereby confirming the existence of bubble doubling transition to SNA. This may be explained as due to the fact that in the bubble doubling transition, the unaffected regions of the strands of the period$-$2 torus remain so even after the birth of SNA.  
\section{\label{sec:level2}Summary and conclusion}
In this paper, we have reported a novel machanism for the birth of strange nonchaotic attractor in quasiperiodically forced Chua's circuit. We term this route as {\em bubble doubling route} to SNA. At first, we have presented the appearence of bubble through Poincar\'e surface of cross section and the bifurcation of the bubble by varing the control parameter $f_{1}$ in the suitable range. Further, we used power spectrum to qualitatively distinguish quasiperiodicity, SNA and choatic attractor. Followed by this, we have computed Lyapunov exponent and its variance as a function of control parameter $f_{1}$ and these clearly distinguish the dynamical region of quasiperiodicity, SNA and chaos. To confirm the presence of SNA we have computed singular continuous spectrum, which clearly shows the fractal dimension of the SNA and its fractal path in the complex plane. The separation of near by point test and distribution of the finite time Lyapunov exponent also explains the presence of SNAs. It is to be noted that the formation of SNA through this bubble doubling mechanism is significantly different from the well known mechanisms in the literature (Table - I).

\end{document}